\documentclass[12pt]{iopart}

\usepackage{iopams} 
\usepackage{bm,graphics,graphicx,epsfig,color}

   \newcommand{\prd}{Phys. Rev. D}

   \newcommand{\aj}{Astron. J.}

\newcommand{\bv}{{\bm v}}
\newcommand{\bx}{{\bm x}}
 
\def\tp{\tilde{p}}
\def\te{\tilde{e}}
\def\tal{\tilde{\alpha}}
\def\tbe{\tilde{\beta}}
\def\tom{\tilde{\omega}}

\newcommand{\llyc}{L$^2$YC}

\begin{document}

\title[Late-time eccentricity]{Compact binary inspiral: Nature is perfectly happy with a circle}

\author{Clifford M. Will$^{1,2}$}

\ead{cmw@phys.ufl.edu}
\address{
$^1$ Department of Physics,
University of Florida, Gainesville FL 32611, USA
 \\
$^2$ GReCO, Institut d'Astrophysique de Paris, UMR 7095-CNRS,
Sorbonne Universit\'e, 98$^{bis}$ Bd. Arago, 75014 Paris, France
}

\begin{abstract}
It is standard lore that gravitational radiation reaction circularizes the orbits of inspiralling binary systems.  But in recent papers, Loutrel et al.\ \cite{2019CQGra..36aLT01L,2019CQGra..36b5004L} have argued that at late times in such inspirals, one measure of eccentricity actually increases, and that this could have observable consequences.  We show that this variable, the magnitude of the Runge-Lenz vector ($e_{\rm RL}$), is {\em not} an appropriate measure of orbital eccentricity, when the eccentricity is smaller than the leading non-Keplerian perturbation of the orbit.     Following Loutrel et al., we use Newtonian equations of motion plus the leading gravitational radiation-reaction terms, the osculating-orbits approach for characterizing binary orbits, and a two-timescale analysis for separating secular from periodic variations of the orbit elements.  We find that $e_{\rm RL}$ does grow at late times, but that the actual orbital variables $r$ and $dr/dt$  show no such growth in oscillations.  This is in complete agreement with Loutrel et al.  We reconcile this apparent contradiction by pointing out that it is essential to take into account the {\em direction} of the Runge-Lenz vector, not just its magnitude.  At late times in an inspiral, that direction, which defines the pericenter angle, advances at the {\em same rate} as the orbital phase.   The correct picture is then of a physically circular orbit whose osculating counterpart is indeed eccentric but that resides permanently at the orbit's {\em latus rectum} at $-90^{\rm o}$, therefore exhibiting no oscillations.   
Including first post-Newtonian effects in the equations of motion, we show that $e_{\rm RL}$ grows even more dramatically.  But the phase of the Runge-Lenz vector again rotates with the orbit at late times, but now the osculating orbit resides at ``perpetual apocenter'', so again the physical orbit circularizes.
\end{abstract}


\maketitle

\section{Introduction}
\label{sec:intro}

Gravitational radiation emits both energy and angular momentum to infinity.  As a result, its long-term effect on binary orbits is to circularize them \cite{1963PhRv..131..435P,1964PhRv..136.1224P}.  On the other hand, it is well-known in celestial mechanics that, while a strictly circular orbit is a perfectly suitable solution in appropriate circumstances, the {\em limiting process} from an eccentric orbit to a circular orbit must be handled with extreme care.  The reason is that orbital eccentricity is, by its very definition, accompanied by another orbit element, the pericenter angle.   This is because an eccentric orbit has an orientation attached to it.   The natural way to define these orbital elements is via the Runge-Lenz vector (sometimes called the Laplace-Runge-Lenz vector), defined by
\begin{equation}
{\bm A} \equiv \frac{{\bm v} \times {\bm h}}{GM} - {\bm n} =  e \cos \omega \,{\bm e}_X  + e \sin \omega \, {\bm e}_Y \,,
\end{equation}
where ${\bm v} \equiv {\bm v}_1 - {\bm v}_2$ is the orbital relative velocity, $\bm n \equiv \bx/r$ is the radial unit vector ($\bx = \bx_1 - \bx_2$, $r = |\bx|$), ${\bm h} \equiv \bx \times \bv$ is the angular momentum per unit reduced mass, $M$ is the total mass of the binary system and $G$ is Newton's constant.    The vector $\bm A$ lies in the orbital plane, and has components $\alpha \equiv e \cos \omega$ and $\beta \equiv e \sin \omega$ relative to a reference basis ${\bm e}_X$ and ${\bm e}_Y$, where $e$ and $\omega$ are the eccentricity and pericenter angle, respectively.

But when $e \to 0$, the vector $\bm A$ becomes null (in the vector algebra, not spacetime sense), and $\omega$ becomes meaningless.  This limit is clearly singular, and therefore must be treated very carefully, lest one draw misleading conclusions.

In two recent papers, Loutrel et al.\ \cite{2019CQGra..36aLT01L,2019CQGra..36b5004L} (hereafter \llyc ) pointed out that, during the course of a compact binary inspiral driven by gravitational radiation reaction, the ``Runge-Lenz'' eccentricity $e_{\rm RL} \equiv | {\bm A} |$ can actually {\em increase} at late times.   They even suggested  that ``nature abhors a circle'' \cite{2019CQGra..36aLT01L}.  On the other hand, they pointed out that other measures of eccentricity are compatible with circularizing orbits, and studied whether different versions of eccentricity were more suitable for encapsulating the late time behavior of inspiraling orbits.   Similar conclusions were reached by Ireland et al.\ \cite{2019arXiv190403443I} in a study of eccentric inspiral with spin couplings.

In this paper, we attempt to bring some clarity to this situation, by pointing out that, in the post-Newtonian limit of the two-body equations of motion, the magnitude of the Runge-Lenz vector is a wholly {\em inappropriate} measure of orbital eccentricity.   On the other hand, when properly treated as a vector, $\bm A$ does provide a valid description of a circularizing orbit in a manner that is compatible with other measures of the orbital properties.

As the ancient philosophers proclaimed, the circle is perfectly fine, in nature's opinion.    

We use the same equations of motion as \llyc\ (Newtonian gravity plus the leading gravitational radiation reaction terms); treat the orbits using the same formalism of osculating orbit elements, and analyze the resulting ``Lagrange planetary equations'' for the evolution of those elements using the same two-timescale method for separating the evolution of the orbit elements into ``secularly'' varying contributions that evolve on a long radiation-reaction timescale, and periodic contributions that vary on an orbital timescale.  We treat the same model inspiral as \llyc, an equal mass binary with initial eccentricity $0.01$ and initial semilatus rectum $p = 20 GM/c^2$.  We find that the Runge-Lenz eccentricity $e_{\rm RL}$ initially decreases (with periodic oscillations superimposed), but reaches a minimum and increases, ending at a value around $0.03$ when the orbit evolution is terminated at $p = 6GM/c^2$.  Our results agree perfectly with \llyc.   We trace this behavior to a specific periodic term in the solution for the $X$ and $Y$ components ($\alpha$ and $\beta$) of the vector $\bm A$.  

On the other hand, when we use the solutions for the orbit elements (including both secular and periodic parts) to reconstruct the actual orbital variables $r$ and $\dot{r}$ as functions of the orbital phase $\phi$, we find that the initial oscillations in $r$ and $\dot{r}$ decrease in amplitude, until the orbit makes a transition to a quasicircular inspiral.  This behavior was also noted by \llyc.  There is no apparent increase in eccentricity in the orbit itself.  Furthermore, the ``orbit averaged'' eccentricity, constructed from the orbit averaged $X$ and $Y$ components  of $\bm A$ also decreases monotonically.  

How can an orbit have its eccentricity increase at the same time as it becomes more circular?
The answer lies in understanding the behavior of the {\em direction} of $\bm A$ at late times.

\begin{figure*}[t]
\begin{center}

\includegraphics[width=3in]{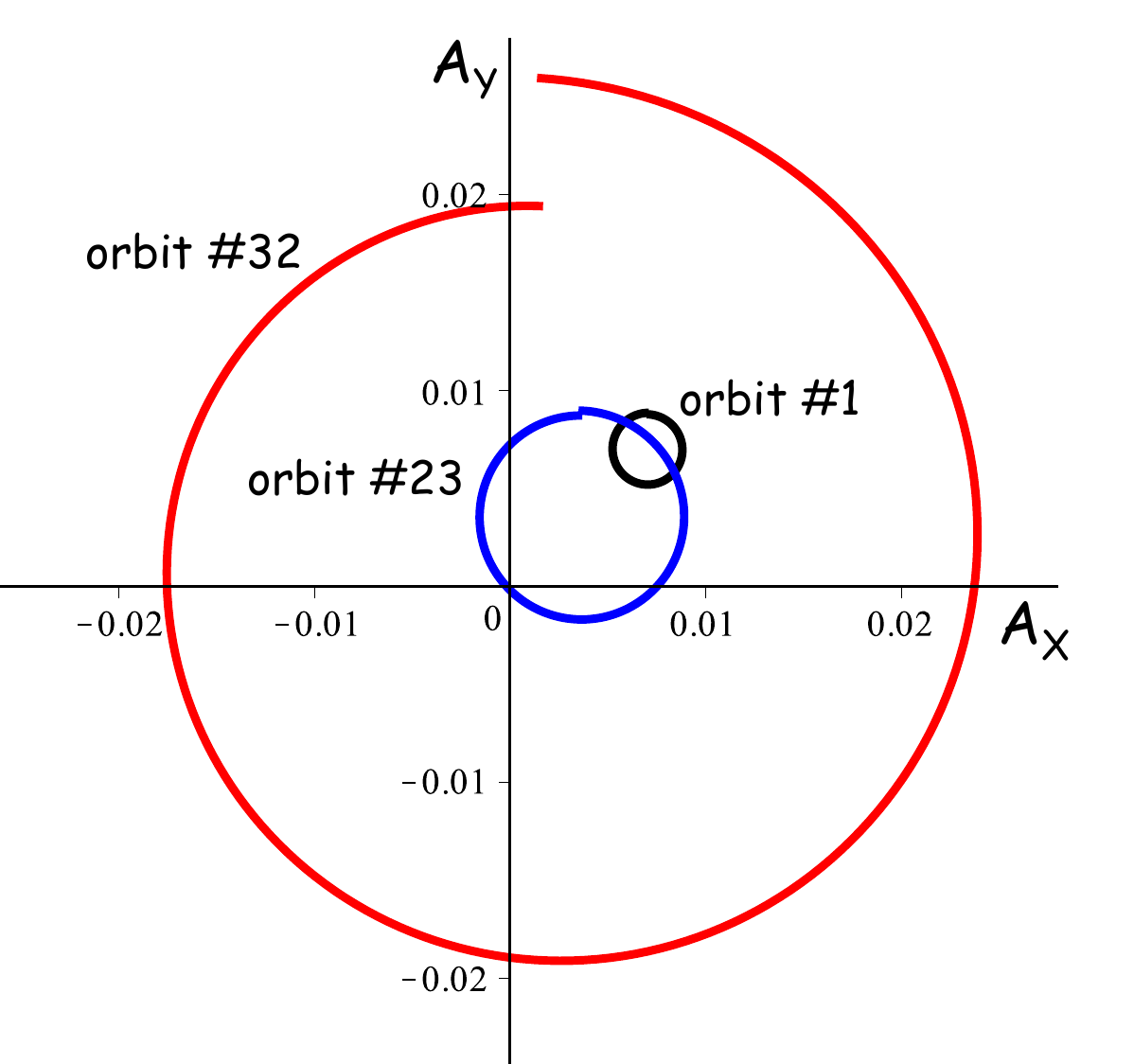}

\caption{\label{fig:RLphase} Evolution of the Runge-Lenz vector in an inspiralling binary.  Initially $\bm A$ points in the direction of the pericenter ($45^{\rm o}$) with a magnitude given by the initial eccentricity ($0.01$).  During the first orbit, the tip of the vector revolves around this point in the $A_X$-$A_Y$ plane (black circle) because of the periodic contributions to $\bm A$.  By orbit \#23, the center of the circle has migrated inward (decreasing $\te$), but the circle has grown; the Runge-Lenz vector even passes near the origin.  By the end of the evolution, orbit \#32, as the orbital phase runs from zero to $2\pi$, the phase of $\bm A$ runs from $\pi/2$ to $5\pi/2$, in lock step with the orbital phase, but offset by $90^{\rm o}$. } 
\end{center}
\end{figure*}

It turns out that the increase of $e_{\rm RL}$ begins when the average eccentricity $\te$ (tildes will denote the ``orbit averaged'' elements) decreases to a point where $\te \sim (64/5)\eta (GM/c^2\tp)^{5/2}$, where $\eta = m_1m_2/M^2$ is the dimensionless reduced mass.  In other words, the growth begins when $\te$ is of the same order as the amplitude of the leading non-Keplerian perturbation, in this case, radiation reaction.   However, at this point the Runge-Lenz {\em vector} changes its behavior dramatically.  In the limit of small eccentricities, it turns out (Sec.\ \ref{sec:effects}) that the leading contributions to the Runge-Lenz vector take the form
\begin{equation}
{\bm A} = \te \left ( \cos \tom \, {\bm e}_X + \sin \tom \, {\bm e}_Y \right )
+ \frac{64}{5} \eta \left ( \frac{GM}{c^2\tp} \right )^{5/2} \left (- \sin \phi \,{\bm e}_X + \cos \phi \,{\bm e}_Y \right ) \,,
\label{eq:RLvector}
\end{equation}
where $\te$, $\tp$ and $\tom$ are the orbit-averaged eccentricity, semilatus rectum and pericenter angle, respectively.
This simple form displays all the features of the evolution of $\bm A$ obtained from the full solution of the planetary equations and displayed in Fig.\  \ref{fig:RLphase} for the model inspiral studied by \llyc.  The figure shows the evolution of $\bm A$ during the first, 23rd and 32nd orbits, the last being just prior to the end of the evolution, when the semilatus rectum $\tp$ reaches the value $6GM/c^2$, corresponding roughly to the innermost stable orbit.

In the early part of the inspiral, when $\te > (64/5)\eta (GM/c^2\tp)^{5/2}$, the Runge-Lenz vector points toward the initial pericenter direction ($45^{\rm o}$ in the example displayed in Fig.\ \ref{fig:RLphase}), with an average length of the initial eccentricity, $\te =0.01$.   The tip of the vector revolves around that point because of the periodic terms in Eq.\ (\ref{eq:RLvector}).  This is represented by the black circle in Fig.\ \ref{fig:RLphase}, and is the expected behavior of $\bm A$.      As time passes and $\te$ decreases, the center of the circle, governed by the first term, moves toward the origin (the average pericenter angle $\tom$ is constant in this example of pure radiation reaction), but the diameter of the circle increases, because $\tp$ is decreasing.  This is represented by the blue circle in Fig.\ \ref{fig:RLphase}. By the time of orbit number 32, $\te \ll (64/5)\eta (GM/c^2\tp)^{5/2}$, the second term in Eq.\ (\ref{eq:RLvector}) dominates, and the direction of  $\bm A$ revolves from $\pi/2$  to $5\pi/2$ as the orbital phase $\phi$ advances from $64\pi$ to $66\pi$, in lock step with the orbit.  Its length is roughly $0.03$ corresponding to the late-time value of $e_{\rm RL}$.

At late times, the pericenter angle of the osculating orbit, defined by the direction of $\bm A$, is advancing at the same rate as the orbit itself, so that the  ``true anomaly'', $f  = \phi - \omega$, which defines the angle between the relative vector $\bx$ and the pericenter, is constant, with a value $\approx - \pi/2$. So the osculating orbit that corresponds to the actual orbit is an eccentric orbit at ``perpetual latus rectum''.    Since orbital variables such as $r$ and $\dot{r}$ depend on sines and cosines of $f$, these variables display no evidence of eccentricity, since $f$ is constant.   The physical orbits are therefore circular (or quasicircular, because they are shrinking) even though $e_{\rm RL}$ is growing.

\begin{figure*}[t]
\begin{center}

\includegraphics[width=4in]{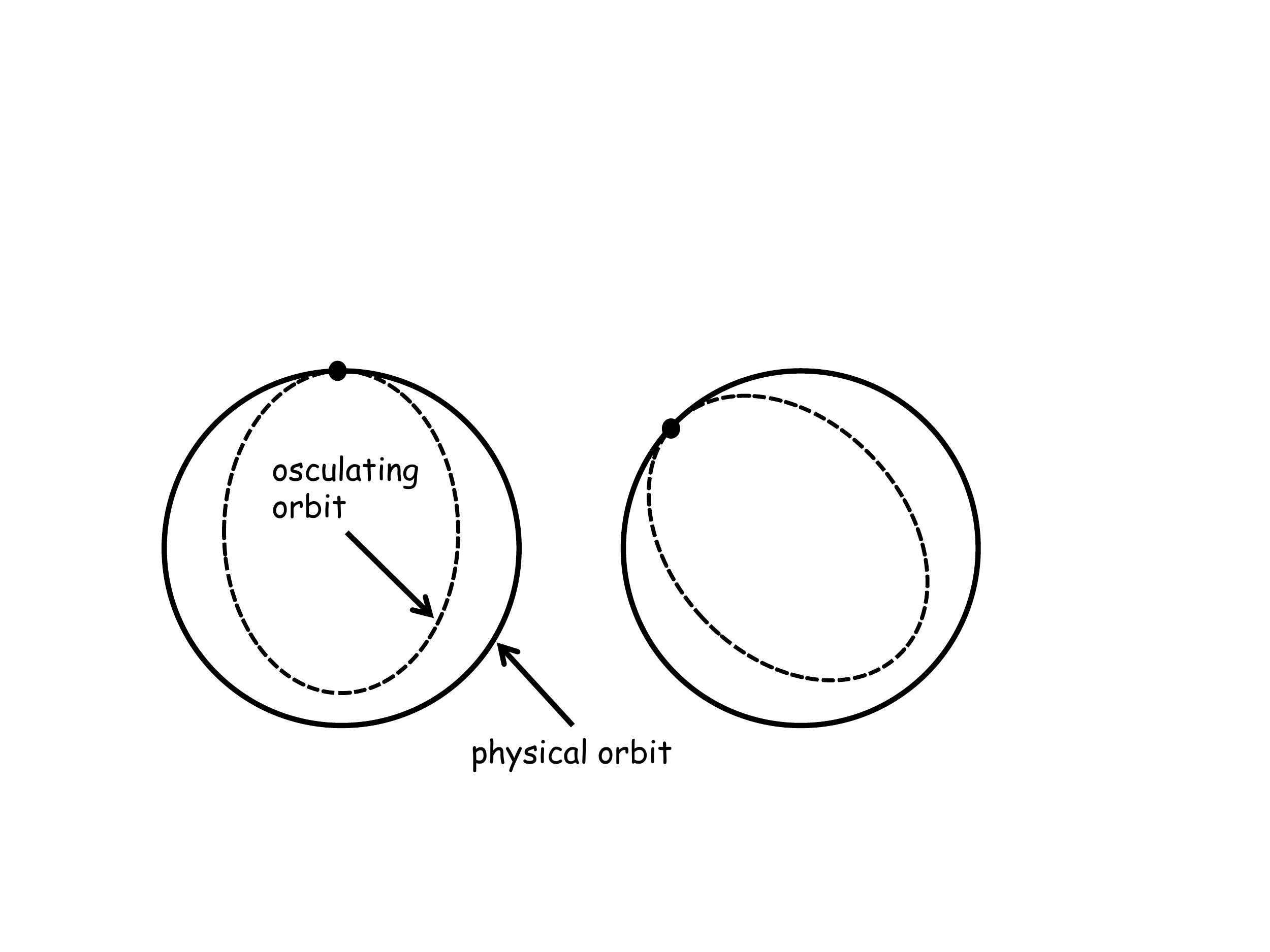}

\caption{\label{fig:orbitfigure} A body moves on a physically circular orbit, while the osculating orbit is eccentric.   The phase of the Runge-Lenz vector (the pericenter angle) advances at the same rate as the body, keeping the osculating orbit at a constant value of  its true anomaly. In the example shown in the figure, the osculating orbit is at ``perpetual apocenter'' (see Sec.\ \ref{sec:1PN}) }
\end{center}
\end{figure*}

This unusual behavior of the Runge-Lenz vector in the small eccentricity limit was first noticed (we believe) by Whitman and Matese \cite{1985CeMec..36...71W}.  Motivated in part by an earlier remark by Greenberg \cite{1981AJ.....86..912G}, who was analysing the precession of rings of Uranus, they pointed out that for a purely radial perturbation of the Kepler problem governed by a force $B(r) {\bm n}$, the osculating orbit corresponding to a circular physical orbit would be an eccentric orbit of perpetual pericenter or apocenter, depending on the sign of $B(r)$.  Lincoln and Will \cite{1990PhRvD..42.1123L} showed that this same phenomenon occurs at the first post-Newtonian order in binary inspiral.  In that case, when $\te$ decreases below a value given by $\approx 3GM/c^2\tp$, the physical orbit is circular while its osculating avatar is an eccentric orbit  of perpetual apocenter, as shown in Fig.\ \ref{fig:orbitfigure}. 

Given the 300-year history of Newtonian celestial mechanics, it may be surprising that this phenomenon is not better known.   In part, the answer is that, for most practical problems of interest to celestial mechanicians, the eccentricity is seldom small enough to matter.  For example, the orbit of Venus is the most circular in the solar system, with $e = 0.0068$, yet that value is 2500 times larger than the scale of the Newtonian perturbation by Jupiter and $10^5$ times larger than the scale of general relativistic effects.   The extremely circular inner binary in the pulsar-triple system J0337+1715 \cite{2014Natur.505..520R} has an eccentricity ($6.9 \times 10^{-4}$) that is over 100 times larger than the scale of the perturbation by the third body and over 400 times larger than the leading GR effect.    
But because gravitational radiation reduces orbital eccentricity while simultaneously increasing the size of relativistic effects, it naturally produces the conditions where these subtleties in treating circular orbits {\em must} be addressed.   

The remainder of this paper provides details and quantitative results.  In Sec.\ \ref{sec:effects} we repeat the calculation of \llyc\ and resolve the apparent contradiction between an increasing $e_{\rm RL}$ and a circularizing physical orbit.  In Sec.\ \ref{sec:1PN} we revisit the Lincoln-Will analysis, adding 1PN corrections in the equations of motion.  We show that the growth of $e_{\rm RL}$ is even more dramatic than in the analysis of \llyc, yet the orbits are still circularized.  In Sec.\ \ref{sec:conclusions} we make concluding remarks.


\section{Late-time eccentricity in an orbit driven by gravitational radiation reaction}
\label{sec:effects}

Following  \llyc,  we use two-body equations of motion that include only the Newtonian acceleration and the leading radiation-reaction contributions, given by
\begin{eqnarray}
{\frac{d{\bm v}}{dt}} &=& - \frac{M{\bm n}}{r^2} + \frac{8}{5} \eta  \frac{ M^2}{ r^3} \left [ \left (3 v^2 + \frac{17}{3} \frac{M}{r} \right )\dot{r} {\bm n}  
- \left ( v^2 + 3 \frac{M}{r} \right ) {\bm v} \right ] \,.
\label{eq:eom}
\end{eqnarray}
In the post-Newtonian (PN) approximation scheme, these are known as 2.5PN terms.
We now use units in which $G=c=1$.

Placing the orbit on the $X$-$Y$ plane, 
we define the ``osculating'' Keplerian orbit using the semilatus rectum $p$, eccentricity $e$,  and pericenter angle $\omega$, defined by the following set of equations:
\begin{eqnarray}
{\bm x} &\equiv& r {\bm n} \,,
\quad 
r \equiv p/(1+e \cos f) \,,
\nonumber \\
{\bm n} &\equiv&  \cos \phi \, {\bm e}_X  +  \sin \phi \, {\bm e}_Y \,, 
\quad
{\bm \lambda} \equiv \partial {\bm n}/\partial \phi \,, 
\quad
\ \hat{\bm h} \equiv {\bm n} \times {\bm \lambda} \,,
\nonumber \\
{\bm h} &\equiv& {\bm x} \times {\bm v} \equiv \sqrt{Mp} \, \bm{e}_Z \,,
\label{eq2:keplerorbit}
\end{eqnarray}
where  $f \equiv \phi - \omega$ is the  {\em true anomaly}, $\phi$ is the orbital phase measured from the $X$-axis and 
 ${\bm e}_A$ are chosen reference basis vectors.   From the given definitions, we see that ${\bm v} = \dot{r} {\bm n} + (h/r) {\bm \lambda}$ and $\dot{r} = (he/p) \sin f$, so that a solution for $r$, $\dot{r}$ and $h$ suffices to determine the orbit $\bx$ and $\bv$. We will work with the  orbit elements $\alpha =e \cos \omega$ and $\beta =  e \sin \omega$;
they are the components of the Runge-Lenz vector
\begin{equation}
{\bm A} \equiv \frac{{\bm v} \times {\bm h}}{M} - {\bm n} = \alpha \,{\bm e}_X  + \beta \, {\bm e}_Y \,,
\end{equation}
which is constant for the pure Keplerian binary orbit.
With these orbit elements we have
\begin{eqnarray}
r &=& p/(1+\alpha \cos \phi + \beta \sin \phi) \,,
\nonumber \\
\dot{r} &=& (h/p) (\alpha \sin \phi - \beta \cos \phi ) \,.
\label{eq2:rdef}
\end{eqnarray}

We then define the radial ${\cal R} \equiv \delta {\bm a} \cdot {\bm n}$ and
cross-track ${\cal S} \equiv \delta {\bm a} \cdot {\bm \lambda}$ components of the perturbing acceleration $\delta {\bm a}$, given by the radiation-reaction terms in Eq.\ (\ref{eq:eom}), and write down the Lagrange planetary equations for the evolution of the orbit elements (see \cite{PW2014} for further discussion),
\begin{eqnarray}
\frac{dp}{d\phi} &=&  \frac{2r^3}{M}  {\cal S}\,,
\nonumber \\
\frac{d\alpha}{d\phi} &=& \frac{r^2}{M} \left [ {\cal R}  \sin \phi   +  {\cal S} (\alpha + \cos \phi) \left ( 1+ \frac{r}{p} \right ) \right ]\,,
\nonumber \\
\frac{d\beta}{d\phi} &=& \frac{r^2}{M} \left [ -{\cal R}  \cos \phi   +  {\cal S} (\beta + \sin \phi) \left ( 1+ \frac{r}{p} \right ) \right ]\,.
\label{eq2:Lagrange}
\end{eqnarray} 

Assuming that the radiation-reaction timescale is suitably long compared to the orbital timescale, we adopt a two-timescale approach
\cite{1978amms.book.....B,1990PhRvD..42.1123L,2004PhRvD..69j4021M,2008PhRvD..78f4028H,2017PhRvD..95f4003W}
for obtaining solutions to the planetary equations.   
Those equations have  the general form
\begin{equation}
\frac{d X_i (\phi)}{d\phi} = \epsilon Q_i (X_j(\phi), \phi) \,,
\label{eq2:dXdf}
\end{equation}
where the subscripts $i$ and  $j$ label the orbit element, and $\epsilon$  is a small parameter that characterizes the perturbation.   
We define the long-timescale variable
$\theta \equiv \epsilon \phi$,
and write the derivative with respect to $\phi$ formally as
$d/d\phi \equiv \epsilon \partial/\partial \theta + \partial/\partial \phi$.
We make an {\em ansatz} for the solution for $X_i (\theta, \phi)$:
\begin{equation}
X_i (\theta, \phi) \equiv \tilde{X}_i (\theta) + \epsilon Y_i (\tilde{X}_j (\theta), \phi) \,.
\label{eq2:ansatz}
\end{equation}
The split is defined such that $\tilde{X}_i=\langle X_i (\theta, \phi) \rangle$ and $\langle Y_i (\tilde{X}_j (\theta), \phi) \rangle = 0$, where the ``average'' 
$\langle \dots \rangle$ is defined by 
$\langle A \rangle \equiv (1/2\pi) \int_0^{2\pi} A(\theta,\phi) d\phi $, holding $\theta$ fixed.  We also define the ``average-free'' part by ${\cal AF}(A) \equiv A(\theta,\phi) - \langle A \rangle $.   Equation (\ref{eq2:dXdf}) then splits into an equation for the long-timescale evolution of the averaged elements, and an equation for the average-free contributions, given by 
\begin{eqnarray}
\frac{d\tilde{X}_i}{d\theta} &=& \langle Q_i (\tilde{X}_j + \epsilon Y_j, \phi) \rangle \,,
\label{eq2:aveq}\\
\frac{\partial Y_i}{\partial \phi} &=& {\cal AF} \left (Q_i (\tilde{X}_j + \epsilon Y_j, \phi) \right )  - \epsilon \frac{\partial Y_i}{\partial \tilde{X}_k} \frac{d\tilde{X}_k}{d\theta} \,.
\label{eq2:avfreeeq}
\end{eqnarray}

It is straightforward to apply this method to the evolution described by Eq.\ (\ref{eq:eom}).
For the evolution of the averaged orbit elements, we find
\begin{eqnarray}
\frac{d\tp}{d\phi} &=& - \frac{8}{5} \eta \tp \left (\frac{M}{\tp} \right )^{5/2} \left (8+7\te^2 \right ) \,,
\nonumber \\
\frac{d\te}{d\phi} &=& - \frac{1}{15} \eta \te \left (\frac{M}{\tp} \right )^{5/2} \left (304+121\te^2 \right ) \,,
\nonumber \\
\frac{d\tom}{d\phi}&=& 0 \,,
\label{eq2:av}
\end{eqnarray}
and for the average-free parts, we find
\begin{eqnarray}
Y_p &=& -\frac{4}{5} \eta \tp  \left (\frac{M}{\tp} \right )^{5/2} \te \,\left ( 36 \sin f + 5\te \sin 2f + 4\te^2 \sin f \right ) \,,
\nonumber \\
Y_\alpha &=& -\frac{1}{180} \eta  \left (\frac{M}{\tp} \right )^{5/2}
\biggl \{  2304 \sin \phi + 1920 \te \sin (2\phi - \tom) 
\nonumber \\
&& \quad
+ 8\te^2 \left [ 576 \sin \phi + 231 \sin (\phi - 2 \tom)
+ 91 \sin (3 \phi - 2\tom) \right ] 
\nonumber \\
&& \quad
+ 15 \te^3 \left [ 10 \sin (2 \phi -3\tom) +62 \sin (2 \phi - \tom) + 7 \sin (4\phi - 3\tom) \right ] 
\nonumber \\
&& \quad
+ 72 \te^4 \left [ 4 \sin \phi + \sin (\phi - 2\tom) + \sin (3\phi - 2\tom) \right ] \biggr \} \,,
\nonumber \\
Y_\beta &=& \frac{1}{180} \eta  \left (\frac{M}{\tp} \right )^{5/2}
\biggl \{  2304 \cos \phi +1920 \te \cos (2\phi - \tom) 
\nonumber \\
&& \quad
+ 8\te^2 \left [ 576 \cos \phi - 231 \cos (\phi - 2 \tom)
+ 91 \cos (3 \phi - 2\tom) \right ] 
\nonumber \\
&& \quad
- 15 \te^3 \left [ 10 \cos (2 \phi -3\tom) - 62 \cos (2 \phi - \tom) - 7 \cos (4\phi - 3\tom) \right ] 
\nonumber \\
&& \quad
+ 72 \te^4 \left [ 4 \cos \phi - \cos (\phi - 2\tom) + \cos (3\phi - 2\tom) \right ] \biggr \} \,,
\label{eq2:Y}
\end{eqnarray}
where $\te^2 = \tal^2 + \tbe^2$ and $\tom = \arctan (\tbe/\tal)$.   Because the claimed eccentricity growth in \llyc\  occurs at order $\eta^2 (M/p)^5$, we have actually carried out the calculation to the next order in $\epsilon$ (see \cite{2017PhRvD..95f4003W} for details).  Note that this is not the same as the next order in radiation-reaction contributions to the equations of motion, which would be $O[\eta (M/\tp)^{7/2}]$.  It turns out that these contributions do not affect our central conclusions..

Recalling that, for each orbit element, $X_i = \tilde{X}_i + Y_i$, we can reconstruct the orbital variables $r$, $\dot{r}$ and $h$, to obtain
\begin{eqnarray}
r &=& \tp\frac{1 - \frac{8}{5} \eta x^{5/2} \te {\cal A}_{2.5} \sin f }{1 + \te \cos f
  -  \frac{1}{180} \eta x^{5/2} \te {\cal B}_{2.5} \sin f  } \,,
  \nonumber \\
\dot{r} &=& x^{1/2} \frac{\te \sin f -  \frac{1}{180} \eta x^{5/2} {\cal C}_{2.5}}{\left ( 1 - \frac{8}{5} \eta x^{5/2} \te {\cal A}_{2.5} \sin f \right )^{1/2}} \,,
  \nonumber \\
h &=& (M \tp )^{1/2} \left (1 - \frac{8}{5} \eta x^{5/2} \te {\cal A}_{2.5} \sin f \right )^{1/2} \,,
\label{eq2:orbit}
\end{eqnarray}
where $x \equiv M/\tp$, and 
\begin{eqnarray}
{\cal A}_{2.5}&=& 18 + 5\te \cos f + 2 \te^2 \,,
\nonumber \\
{\cal B}_{2.5} &=& 1920+5152 \te \cos f +1185 \te^2 +510 \te^2 \cos 2f + 288 \te^3 \cos f \,,
\nonumber \\
{\cal C}_{2.5} &=& 2304 + 1920 \te \cos f + 32 \te^2 \left (144-35 \cos 2f \right ) 
\nonumber \\
&& \quad +15 \te^3 \left ( 62 \cos f - 3 \cos 3f \right ) + 288 \te^4 \,.
\end{eqnarray}
From the three variables $r$, $\dot{r}$, and $h$, one can construct $\bx$ and $\bv$, can link $\phi$ to time via $d\phi/dt = h/r^2$, and can construct the gravitational waveform, using the quadrupole formula $h^{jk} = 4\eta (M/R) (v^j v^k - mn^jn^k/r)$, for example.

Orbital eccentricity is not a unique concept, and many different definitions can be formulated.  Several possibilities were discussed by \llyc.   One is the orbit averaged eccentricity, $\te$.  Another  is the magnitude of the Runge-Lenz vector, defined by
\begin{eqnarray}
e_{\rm RL} &\equiv& \left ( \alpha^2 + \beta^2 \right )^{1/2} 
\nonumber \\
&=& \left [ (\tal + Y_\alpha)^2 + (\tbe + Y_\beta )^2 \right ]^{1/2} \,.
\label{eq2:eRLdef}
\end{eqnarray}
Another definition is denoted the ``Keplerian'' eccentricity by \llyc, given by
\begin{equation}
e_{\rm K} \equiv \frac{ r_{\rm max} - r_{\rm min}}{r_{\rm max} + r_{\rm min}} \,,
\label{eq2:ekepler}
\end{equation}
where $r_{\rm max}$ and $r_{\rm min}$ are the orbital separations at adjacent turning points of the orbit, where $dr/dt = 0$.  We wish to compare the evolution of these quantities with each other and with the actual orbital behavior as described in Eqs.\ (\ref{eq2:orbit}) for an inspiralling binary system.

For small eccentricities, the secular evolution of the averaged elements $\tp$ and $\te$ can be solved straightforwardly from Eqs.\ (\ref{eq2:av}), yielding
$\tp = \tp_0 Z^{2/5} $\,,
and 
$\te = \te_0 Z ^{19/30}$,
where $Z = 1 - 32 \eta ({M}/{\tp_0})^{5/2} \phi $, and $\tp_0$ and $\te_0$ are the initial values at $\phi=0$.  Since we are ignoring lower-order post-Newtonian effects, the average pericenter angle $\tom$ does not change.  Clearly the averaged eccentricity $\te$ decreases monotonically.  

\begin{figure*}[t]
\begin{center}

\includegraphics[width=4in]{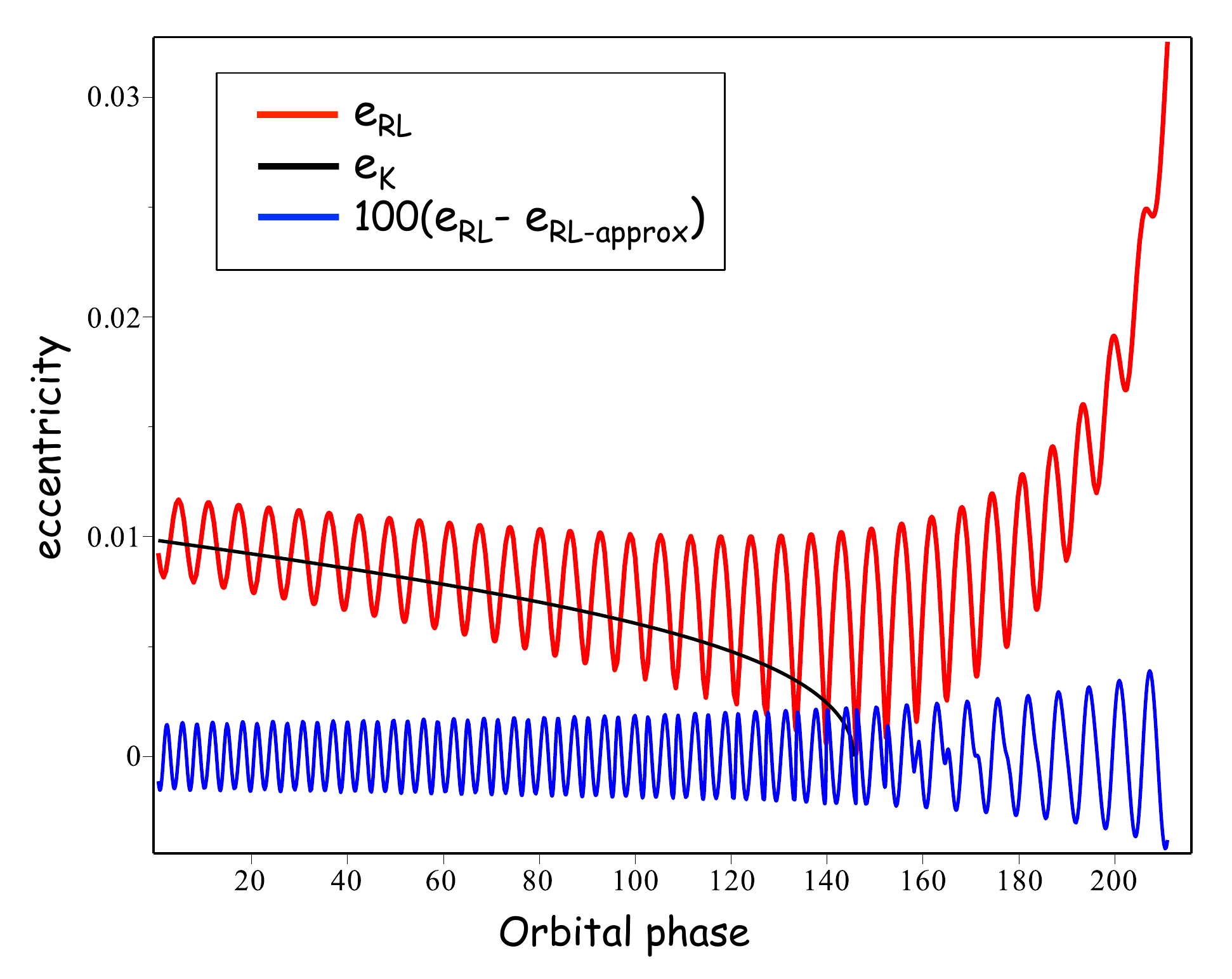}

\caption{\label{fig:eRL} Eccentricities as a function of orbital phase.  In red is the Runge-Lenz eccentricity $e_{\rm RL}$,  while in black is the ``Keplerian'' eccentricity $e_{\rm K}$.  The blue curve is $100 \times$ the difference between $e_{\rm RL}$ using the full solution for the orbit elements and that using the approximate formula Eq.\ (\ref{eq2:eRL}).  Notice that $e_{\rm RL}$ begins to increase and $e_{\rm K}$ vanishes when $\phi = 145$, which is precisely where $\dot{r}$ turns monotonically negative.}
\end{center}
\end{figure*}

To study the behavior of the Runge-Lenz eccentricity, we treat a specific example.  We follow \llyc\ by choosing
an equal mass system ($\eta = 1/4$), with $\te_0 = 0.01$ and $\tp_0 = 20 M$, or $x_0 = 1/20$.   We substitute Eqs.\ (\ref{eq2:Y}) into Eq.\ (\ref{eq2:eRLdef}), and incorporate the evolutions of $\tp$, $\tal$ and $\tbe$ with $\phi$.  The resulting evolution of $e_{\rm RL}$ is shown (in red in the online version) in Fig.\ \ref{fig:eRL}.  The Runge-Lenz eccentricity oscillates because of the contributions of $Y_\alpha$ and $Y_\beta$, but decreases on average, until around $\phi \approx 145$, or around 23 orbits later, when it begins to increase, reaching a value three times larger than the initial value by end of the evolution, when $r \approx 6M$.    This behavior exactly matches that shown in Fig.\ 2 of \cite{2019CQGra..36b5004L}, apart from the fact that we use orbital phase while \llyc\ used time for the evolutions (note that, while the label on Fig.\ 2 of 
\cite{2019CQGra..36b5004L} indicates $e_{\rm K}$, what is actually plotted is $e_{\rm RL}$).  Since the initial eccentricity is so small, this behavior can be well approximated by setting $\te=0$ in  $Y_\alpha$ and $Y_\beta$ in Eq.\ (\ref{eq2:Y}), and including only the leading term in the latter two expressions.  In \cite{2019CQGra..36b5004L}, these contributions come from the leading term with coefficient $-768$ in $C_\alpha^1$ and $S_\beta^1$ in their Eqs.\ (A.11) and (A.30).   With this approximation we can write
\begin{equation}
e_{\rm RL} = \left ( \te^2 - 2 \te {\cal Q} \sin \phi +  {\cal Q}^2  \right )^{1/2} \,,
\label{eq2:eRL}
\end{equation}
where ${\cal Q} \equiv (64/5) \eta (M/\tp)^{5/2}$.   Figure  \ref{fig:eRL} also displays the difference between the approximate expression (\ref{eq2:eRL}) and the full expression for $e_{\rm RL}$, multiplied by $100$.  So we see that all the eccentricity growth comes from the leading terms in  $Y_\alpha$ and $Y_\beta$.  These terms do not vanish as $\te \to 0$.  

On the other hand, if we use this leading behavior and substitute $\alpha = \tal -  {\cal Q} \sin \phi$, $\beta = \tbe +  {\cal Q} \cos \phi$ and $p=\tp$ into the expressions for $r$ and $\dot{r}$ in Eqs.\ (\ref{eq2:rdef}), we find that
\begin{eqnarray}
r &=& \frac{\tp}{1+ \tal \cos \phi + \tbe \sin \phi} \,,
\nonumber \\
\dot{r} &=& \left ( \frac{M}{\tp} \right )^{1/2} \left ( \te \sin \phi -  {\cal Q} \right )\,.
\label{eq2:orbit2}
\end{eqnarray}
The leading terms in $Y_\alpha$ and $Y_\beta$ that contribute to the growth of $e_{\rm RL}$ actually {\em cancel} in the orbital variable $r$.  Thus the orbit becomes highly circular as $\te \to 0$, apart from  residual oscillations coming from the smaller $\te$-dependent terms in $Y_p$, $Y_\alpha$ and $Y_\beta$, whose amplitude never exceeds a few parts in $10^5$ for the model inspiral. 

Those terms do not cancel in $\dot{r}$, but instead of generating apparent eccentricity, they produce the $\phi$-independent term $-{\cal Q}$, which enables the smooth transition of $\dot{r}$ from an oscillatory behavior to a monotonic decrease when $\te$ decreases and $\cal Q$ increases sufficiently.   

\begin{figure*}[t]
\begin{center}

\includegraphics[width=5in]{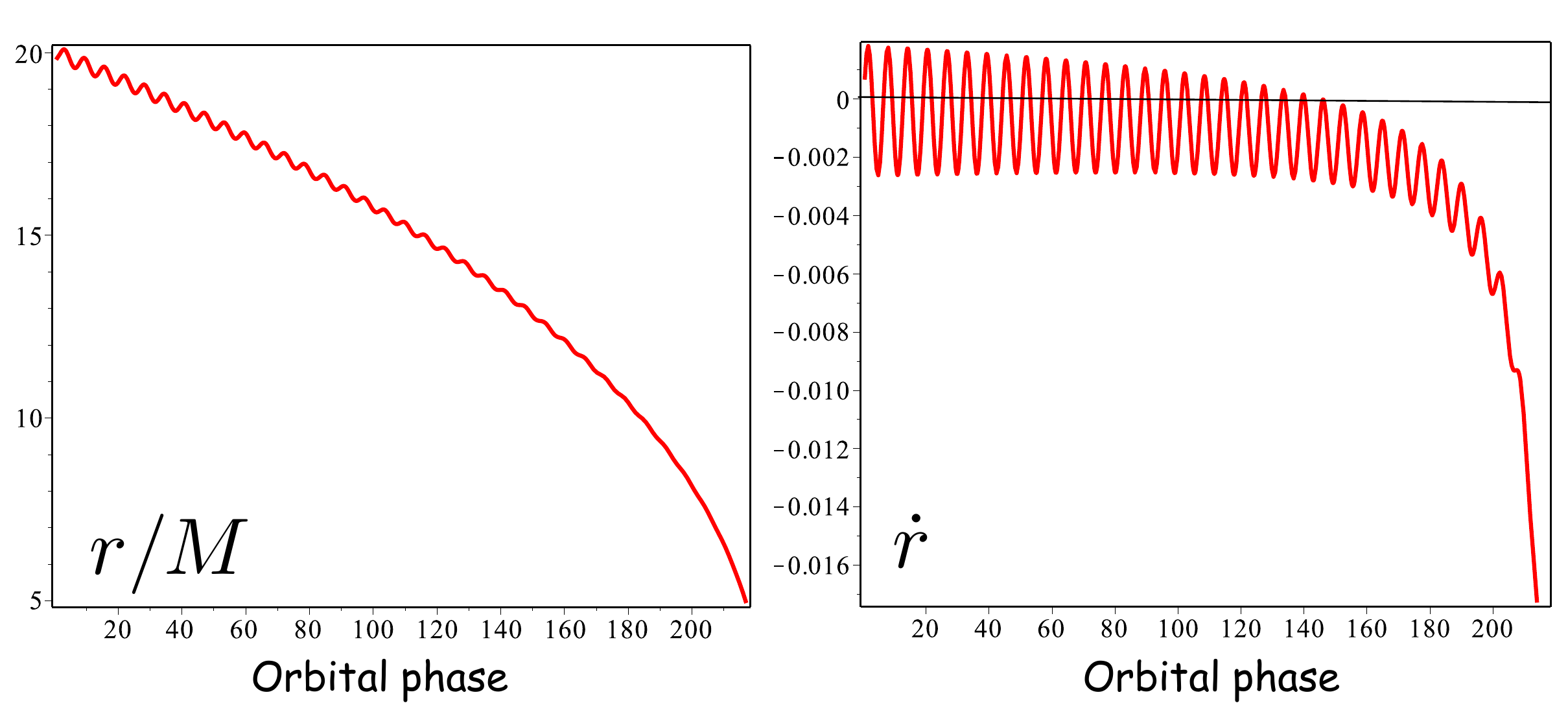}

\caption{\label{fig:rrdotRR} Orbital separation $r$ and radial velocity $\dot{r}$ as functions of orbital phase $\phi$.  The oscillations in $r$ and $\dot{r}$ damp out, consistent with a decreasing orbital eccentricity.  After $\phi =145$, $\dot{r}$ is consistently negative, implying a plunge with no further turning points of the orbit. }
\end{center}
\end{figure*}

In Figure \ref{fig:rrdotRR}, we illustrate this by plotting the full solutions for $r$ and $\dot{r}$ as functions of $\phi$, including {\em all} the contributions to $Y_p$, $Y_\alpha$ and $Y_\beta$ through $O({\cal Q}^2)$.  We again use the inspiral example of \llyc.   The orbital separation shows oscillations of decreasing amplitude followed by a monotonic decrease.  The radial velocity also oscillates, but it becomes strictly negative when $\te = {\cal Q}$.  For the model inspiral, this occurs when $\phi = 145.6$.    Notice that this is {\em exactly} the place where the Runge-Lenz eccentricity begins its apparent increase. 

The Keplerian eccentricity does not show any increase at late times.  We can use Eqs.\ (\ref{eq2:orbit}), and work in the small $\te$ limit, to show that turning points of the orbit occur when $\te \sin f =  {\cal Q}$, or where $\cos f = \pm [1-({\cal Q}/\te)^2]^{1/2}$.   Substituting these values into the expression for $r$ and using Eq.\ (\ref{eq2:ekepler}), we obtain
\begin{equation}
e_{\rm K} = \te \frac{\sqrt{1-({\cal Q}/\te)^2}}{1- 5{\cal Q}^2/6} \,.
\end{equation}
This eccentricity decreases monotonically, reaching zero at $\phi = 145.6$ ($\te = {\cal Q}$) where there are no more turning points and $e_{\rm K}$ becomes meaningless.  This eccentricity is plotted in black in Fig.\ \ref{fig:eRL}.    

We reconcile the growth of $e_{\rm RL}$ with the evident circularization of the physical orbit by studying the actual Runge-Lenz vector.  Inserting the leading contributions to $Y_\alpha$ and $Y_\beta$ into the formula ${\bm A} = \alpha {\bm e}_X + \beta {\bm e}_Y$, we find
\begin{equation}
{\bm A} = \te \left ( \cos \tom \, {\bm e}_X + \sin \tom \, {\bm e}_Y \right )
+  {\cal Q} \left (- \sin \phi \,{\bm e}_X + \cos \phi \,{\bm e}_Y \right ) \,.
\label{eq:RLvector2}
\end{equation}
This vector evolves from one that points in the direction of the averaged pericenter angle $\tom$ when $\te > {\cal Q}$ to one that 
 rotates in step with the orbit, but 90 degrees out of phase (see the discussion in Sec.\ \ref{sec:intro}). In other words, the pericenter angle begins to rotate as $\omega = \pi/2 + \phi$.   Thus the orbit separation and radial velocity at late times behave as 
\begin{eqnarray}
r ^{-1}&=& \tp^{-1} \left \{1+ e_{\rm RL} \cos (\phi - \omega) \right \}
 \to  \tp^{-1} \left \{ 1+ {\cal Q} \cos (-\pi/2) \right \} =\tp^{-1} \,,
\nonumber \\
\dot{r} &=& \left ( \frac{M}{\tp} \right )^{1/2} e_{\rm RL} \sin (\phi - \omega)
\to - {\cal Q}\left ( \frac{M}{\tp} \right )^{1/2}    \,,
\end{eqnarray}
in complete agreement with the reconstructed orbit variables $r$ and $\dot{r}$ in Eqs.\ (\ref{eq2:orbit2}) in the limit $\te \to 0$.    


\section{The effects of 1PN corrections}
\label{sec:1PN}

We now add the 1PN corrections to the equations of motion (\ref{eq:RLvector}).   These terms have the form 
\begin{eqnarray}
{\bm a}_{\rm PN} &=& -  \frac{M}{ r^2} \left [  \left ( v^2 - 4\frac{M}{r} \right ){\bm n} - 4 \dot{r} {\bm v} \right ]  \,.
\end{eqnarray}
Carrying out the two-timescale analysis of the Lagrange planetary equations with these 1PN terms we find $d\tp/d\phi = d\te/d\phi = 0$, and $d\tom/d\phi = 3M/\tp$ and 
\begin{eqnarray}
Y_p^{\rm PN} &=& - 4 {M} \te (2-\eta) \cos (\phi - \tom) \,,
\nonumber \\
Y_\alpha^{\rm PN} &=& -  \frac{M}{\tp} \biggl \{  (3-\eta) \cos \phi 
      + \frac{1}{2} \te (5 - 4\eta) \cos (2\phi - \tom) 
 \nonumber \\
 &&
 + \frac{\te^2}{8} \left [ 2(12 - 17\eta) \cos \phi + (32-13\eta) \cos (\phi - 2 \tom ) 
 \right .
 \nonumber \\
 && 
 \left . \quad
 - \eta \cos (3\phi - 2\tom) \right ] \biggr \} \,,
\nonumber \\
Y_\beta^{\rm PN} &=&    -  \frac{M}{\tp} \biggl \{  (3-\eta) \sin \phi 
      + \frac{1}{2} \te (5 - 4\eta) \sin (2\phi - \tom) 
 \nonumber \\
 &&
 + \frac{\te^2}{8} \left [ 2(12 - 17\eta) \sin \phi - (32-13\eta) \sin (\phi - 2 \tom ) 
 \right .
 \nonumber \\
 && 
 \left . \quad
 - \eta \sin (3\phi - 2\tom) \right ] \biggr \} \,,    
\end{eqnarray}
Reconstructing the orbit, we find
\begin{eqnarray}
r &=& \tp\frac{1 - x{\cal A}_1- \frac{8}{5} \eta x^{5/2} \te {\cal A}_{2.5} \sin f }{1 + \te \cos f - x{\cal B}_1 
  -  \frac{1}{180} \eta x^{5/2} \te {\cal B}_{2.5} \sin f  } \,,
  \nonumber \\
\dot{r} &=& x^{1/2} \frac{\te \sin f (1+\frac{1}{2}x{\cal C}_1)-  \frac{1}{180} \eta x^{5/2} {\cal C}_{2.5}}{\left ( 1 - x{\cal A}_1- \frac{8}{5} \eta x^{5/2} \te {\cal A}_{2.5} \sin f \right )^{1/2}} \,,
  \nonumber \\
h &=& (M \tp )^{1/2} \left (1 - x{\cal A}_1 -\frac{8}{5} \eta x^{5/2} \te {\cal A}_{2.5} \sin f \right )^{1/2} \,,
\label{eq2:orbitPN}
\end{eqnarray}
where
\begin{eqnarray}
{\cal A}_1 &=& 4 \te (2 - \eta) \cos f \,,
\nonumber \\
{\cal B}_1 &=& 3 - \eta + \frac{1}{2} \te (5-4\eta) \cos f + \frac{\te^2}{4} \left [ (12 - 17 \eta) + (16-7\eta) \cos 2f \right ] \,,
\nonumber \\
{\cal C}_1 &=& 5 - 4\eta - 2\te (8-3\eta) \cos f \,.
\end{eqnarray}

Focusing on the small $\te$ regime, thus keeping only the leading terms in $Y_\alpha$ and $Y_\beta$, we find that $e_{\rm RL}$ can be approximated by
\begin{equation}
e_{\rm RL} = \left ( \te^2 - 2 \te {\cal P} \sin \phi +  {\cal P}^2  \right )^{1/2} \,,
\label{eq2:eRLPN}
\end{equation}
where ${\cal P} = (3-\eta)M/\tp$, and where we have ignored the smaller 2.5PN contributions of order $\eta (M/\tp)^{5/2}$.  At the same time, the Runge-Lenz vector can be approximated by
\begin{equation}
{\bm A} = \te \left ( \cos \tom \, {\bm e}_X + \sin \tom \, {\bm e}_Y \right )
-  {\cal P} \left ( \cos \phi \,{\bm e}_X + \sin \phi \,{\bm e}_Y \right ) \,.
\label{eq:RLvector3}
\end{equation}
Here again, $e_{\rm RL}$ decreases initially as $\te$ decreases, but then grows as $\cal P$ when the decreasing $\te$ is smaller than the growing $\cal P$.  But the Runge-Lenz vector undergoes the same behavior as we saw in Sec.\ \ref{sec:effects}, pointing initially toward the average pericenter angle $\tom$, but finally revolving around the origin with length $\cal P$ and phase $\phi + \pi$, in lock step with the orbit.  In this case the osculating orbit is one of perpetual apocenter, as first pointed out by Lincoln and Will \cite{1990PhRvD..42.1123L}.   

\begin{figure*}[t]
\begin{center}

\includegraphics[width=3in]{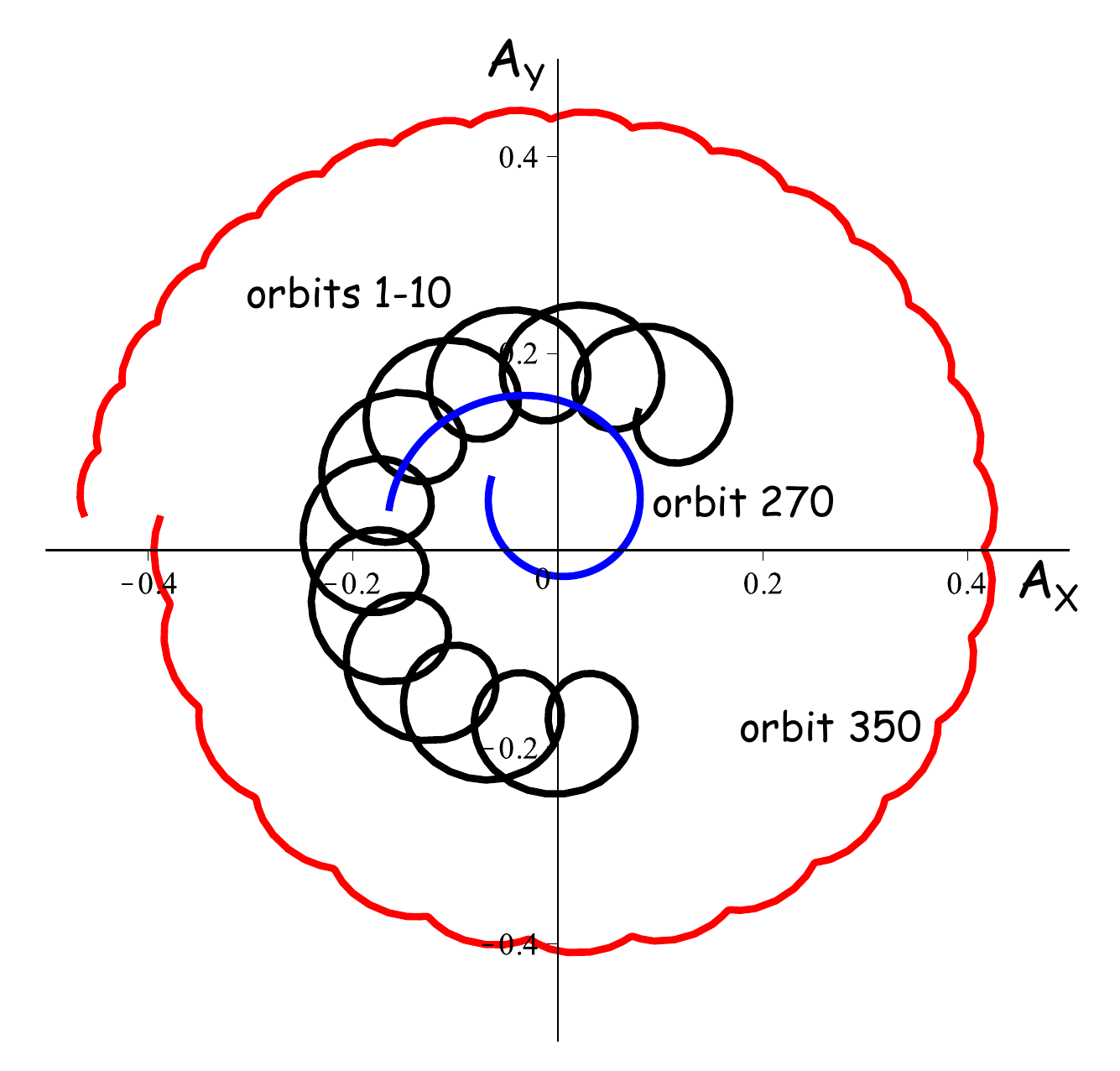}

\caption{\label{fig:RLphase2} Evolution of the Runge-Lenz vector with 1PN corrections.  Initially $\bm A$ points in the direction of the pericenter ($45^{\rm o}$) with a magnitude given by the initial eccentricity ($0.2$).  During the initial orbits, the tip of the vector traces a rosette pattern caused by the periodic terms combined with  the average pericenter advance at $6\pi M/\tp$ radians per orbit (black).    By orbit \#270, the center of the circle has migrated inward (decreasing $\te$), but the circle has grown (blue); the Runge-Lenz vector even passes near the origin.  By orbit \#350, near the end of the evolution (red),  as the orbital phase runs from zero to $2\pi$, the phase of $\bm A$ runs from $\pi$ to $3\pi$, in almost lock step with the orbital phase, but offset by $180^{\rm o}$. } 
\end{center}
\end{figure*}

For the example inspiral discussed by \llyc, with $\te_0 = 0.01$ and $\tp_0 = 20 M$, the physical orbital variables $r$ and $\dot{r}$ behave qualitatively as shown in Fig.\ \ref{fig:rrdotRR}, with a circularizing orbit followed by a plunge.   But in that example, $\te_0$ is already well below $(3-\eta)M/\tp_0$, and so the Runge-Lenz vector is already rotating with the orbit, while $e_{\rm RL}$ is already in its increasing phase.   To see the transition from a conventional Runge-Lenz vector to a rotating vector, we choose initial conditions $\te_0 = 0.2$ and  $\tp_0 = 50 M$, with results shown in 
Fig.\ \ref{fig:RLphase2}.   Notice that, when 1PN terms are included, the transition in the behavior of $\bm A$ occurs much earlier in the inspiral than in Sec. \ref{sec:effects}, simply because the 1PN perturbations are larger than the radiation reaction perturbations.

\section{Conclusions}
\label{sec:conclusions}

We have studied the late-time growth of an orbital ``eccentricity'' defined by the norm of the Runge-Lenz vector during the inspiral of a compact binary system under gravitational radiation reaction, with results in agreement with those of Loutrel et al. \cite{2019CQGra..36aLT01L,2019CQGra..36b5004L}.  But despite this increase, the physical orbit defined by $r$ and $\dot{r}$ circularizes as expected.   

We have resolved this apparent contradiction by pointing out that, when the orbit-averaged osculating eccentricity drops below a value defined by the size of the leading non-Keplerian perturbation,  the direction of the Runge-Lenz vector changes from pointing toward a fixed or slowly revolving orbital pericenter, to pointing toward a direction that rotates in lock-step with the orbit itself.   The result is that the physical orbit may be highly circular, but the osculating orbit to which it is tangent is eccentric, but with the angle between the orbit and the pericenter of its osculating counterpart remaining fixed.   When the dominant perturbation is of 1PN order, the osculating orbit is at perpetual apocenter; when the dominant perturbation is of radiation-reaction order, the osculating orbit is at perpetual latus-rectum.  

Because the physical orbit displays no evidence of growing eccentricity, and since the variables that describe that orbit, $r$, $\dot{r}$ and $h$ are sufficient to calculate a gravitational waveform, there is no reason to expect anomalies in the gravitational waveforms of binary inspiral arising from the growth of $e_{\rm RL}$.   

It is important to point out that the growth of $e_{\rm RL}$ and the unusual phenomena surrounding it have nothing to do with strong-field gravity, or the validity of the post-Newtonian approximation.   These issues can arise just as easily in a purely Newtonian context, if the eccentricity is small enough; indeed the early discussions of strange behavior of the Runge-Lenz vector were motivated by the rings of Neptune  \cite{1985CeMec..36...71W,1981AJ.....86..912G}.   The issue arises in binary inspiral because gravitational radiation reaction naturally produces the right conditions.   The growth effect also has nothing to do with the use of osculating orbit elements and a two-timescale analysis.   Rather, the growth effect is purely a misleading artefact of using $e_{\rm RL}$ as a proxy for eccentricity.  Once one recognizes that, in the circular limit, one must  pay attention to the direction of the Runge-Lenz vector, and not just its magnitude, the mystery is solved. 

Thus, there may be similar issues in the heirarchical three-body problem, when the eccentricity of either the inner or outer system is small; this will be a subject of future research.  

We hasten to acknowledge that the essential technical details presented here are in complete agreement with those presented by \llyc\, in their detailed paper \cite{2019CQGra..36b5004L}.   Our purpose here has been to provide some clarity to the problem that they presented.   Gravitational radiation reaction does lead to circularized orbits.

 \ack
This work was supported in part by the National Science Foundation,
Grant No.\ PHY 16--00188.   We are particularly grateful to Nicholas Loutrel for useful comments on an earlier draft of this paper.     



\section*{References}

\bibliographystyle{iopart-num}

\providecommand{\newblock}{}

\end{document}